# Intensity Normalization Techniques and Their Effect on the Robustness and Predictive Power of Breast MRI Radiomics

Florian Schwarzhans[1], Geevarghese George[1], Lorena Escudero Sanchez[2,3], Olgica Zaric[1], Jean E Abraham[3,4], Ramona Woitek[1,2,+], Sepideh Hatamikia[1,5,+]

1 Research Center for Medical Image Analysis and Artificial Intelligence (MIAAI), Department of Medicine, Danube Private University, Krems, Austria, Rathausplatz 1, AT-3500 Krems-Stein
2 Department of Radiology, University of Cambridge, UK
3 Cancer Research UK Cambridge Centre, University of Cambridge, Li Ka Shing Centre, Robinson Way, Cambridge, CB2 0RE
4 Precision Breast Cancer Institute, Department of Oncology, University of Cambridge, Cambridge, UK, CB2 0QQ
5 Austrian Center for Medical Innovation and Technology (ACMIT), Wiener Neustadt, Austria, Viktor Kaplan-Straße 2/1, 2700 Wiener Neustadt

## Abstract

Radiomics analysis has emerged as a promising approach for extracting quantitative features from medical images to aid in cancer diagnosis and treatment. However, radiomics research currently lacks standardization, and radiomics features can be highly dependent on the acquisition and pre-processing techniques used. In this study, we aim to investigate the effect of various intensity normalization techniques on the robustness of radiomics features extracted from MRI scans of breast cancer patients. The images used are from the publicly available multicenter Investigation of Serial Studies to Predict Your Therapeutic Response with Imaging and molecular Analysis (I-SPY TRIAL) dataset [1], which contains MRI scans of stage 2 or 3 breast cancer patients and from the Platinum and PARP inhibitor for Neoadjuvant treatment of Triple Negative and / or BRCA positive breast cancer (PARTNER) trial [2]. We compared the effect of commonly used intensity normalization techniques, including N4 bias field correction, z-score normalization and piecewise linear histogram matching on the robustness of radiomics features using Intraclass Correlation Coefficient (ICC) between multiple combinations of normalization approaches, identified categories that are robust and therefore could be compared between studies regardless of the pre-processing used. We were able to show that while systematic differences between MRI scanners can significantly affect many radiomics features, a combination of Bias Field correction with piecewise linear histogram normalization can mitigate some of the effects compared to other normalization methods investigated in this paper. Additionally, we observed that the robustness of features is not solely dependent on the normalization method used, but also influenced by the data itself. For prediction of complete pathologic response, we observed that either z-score normalization, or piecewise linear histogram normalization – both in combination with Bias field correction – yielded the best performance. Overall, we were able to demonstrate the importance of carefully selecting and standardizing normalization methods for accurate and reliable radiomics analysis in breast MRI scans.

# 1. Introduction

Radiomics-based image analysis is a rapidly evolving field that aims to extract quantitative features from medical images, to improve disease diagnosis, prognosis and treatment planning [3]. The extracted features are of various classes, including shape-based, first-order, second-order and higher-order statistical features [4]. They provide a quantitative description of the geometrical characteristics of the region of interest (ROI) and can be influenced by factors such as image acquisition, reconstruction settings, ROI segmentation and image pre-processing [5]. As a consequence, radiomics analyses are highly dependent on the normalization methods used to pre-process the images[4]. Multiple factors influence the feature values, which can be attributed to differences in either image acquisition and reconstruction, such as different scanners, ROI segmentation and image pre-processing, or biological variability due to patient inherent factors. To improve the robustness of radiomics studies, several solutions have been proposed, such as eliminating unstable features, correcting for influencing factors, harmonizing datasets and harmonizing extracted radiomics features using the ComBat method [5]–[9].

Magnetic Resonance Imaging (MRI) is a widely used medical imaging modality that is known for its excellent soft tissue contrast and lack of ionizing radiation. While MRI has become an invaluable tool in the diagnosis and monitoring of various diseases, including cancer [1], [10], [11], the interpretation and quantitative analysis of MRI data can be challenging since MRI intensities are non-standardized and highly dependent on the hardware characteristics, sequence type and acquisition parameters [12]–[15]. Consequently, large variability in image intensities in inter-patient and intra-patient comparisons exists that could highly affect the analysis of radiomics features, compromising the reproducibility of measurements [12].

Pre-processing is an essential step in the analysis of medical images, particularly in MRI, where the intensity values of voxels do not have an absolute meaning unlike Computed Tomography (CT) and can vary between scans. The normalization of MRI images is a crucial pre-processing step that aims to reduce variability in image intensity values, making it possible to compare images acquired from different scanners. However, there is currently no standard normalization method used in MRI, leading to challenges in comparing results across different studies and research centers. This is especially problematic when developing radiomics models, as radiomics features have been shown to have a high sensitivity to variations in input images which can arise due to using different normalization techniques, or systematic differences due to the use of different scanners [16].

To solve this problem, previous radiomics studies have focused on image pre-processing techniques. For example, it has been shown that bias field correction efficiently minimizes MR intensity inhomogeneity within a tissue region. The variability generated by different voxel sizes can also be reduced by spatial resampling [17]–[19]. Moreover, ROI definition in terms of the relevant region within the scan (e.g. brain extraction to remove the skull in brain MRI scans) to define the regions in which intensities should be considered before performing normalization can be very beneficial [20]. However, even though these types of pre-processing of MRI scans are widely accepted by the community, there is no consensus within radiomics studies regarding the applied image normalization method [13], [20].

Several studies have investigated the effect of MRI-normalization techniques on the robustness of radiomics features. Carré et al. [20] investigated three normalization methods deemed representative within current radiomics studies (Nyul [21], [22], WhiteStripe [10] and Z-Score [23]), in combination with altering gray-level discretization based on a tumor classification task. They found that first-order features were significantly influenced by intensity normalization methods – recommending z-score normalization whenever possible, but for texture features the discretization of intensity (using specific bin counts) was more relevant than intensity normalization. Li et. al. [24] investigated how pre-processing methods, like bias field correction and rescaling, and intensity normalization methods helped to reduce scanner effects and to improve the reproducibility in brain MRI radiomics. They found that while intensity normalization approaches (with and without) bias field correction seemingly reduced differences in the images between scanners, feature reproducibility, however, did not significantly improve without an additional ComBat normalization on the extracted features. Whilst methods like Z-Score and Nyul normalization are applicable to all types of MRI images, the other methods used (WhiteStripe, FCM [25], GMM [25] and KDE [25]) are only applicable to brain MRI scans, since white matter is used as a reference for the normalization. In their meta-analysis, Panic et. al. [26] investigated MRI-normalization strategies used for cancer classification studies from 2015 onwards, concluding that apart from custom intensity normalization methods, z-score normalization was the method most often used. Additionally, 65% of papers investigated, used some form of spatial normalization by resizing the volume to a pre-defined new voxel spacing. Whilst several studies have investigated the impact of different normalization techniques on radiomics feature robustness and classification capability [18]–[20], [24], [26]–[28], most studies focused on brain MRI scans and also did not further evaluate which radiomics features may be stable across different normalization methods and datasets, and thus good candidates for classification tasks.

In this study, we performed a comprehensive investigation of the effects of various pre-processing methods including bias field correction, spatial and multiple variants of intensity normalization on the extracted radiomics features in breast MRI scans. We identified radiomics features that are robust against such variations in pre-processing, both for comparisons of two specific pre-processing methods and over all methods. In addition, we thoroughly examined the systematic differences between scanners as manifested in radiomics features to determine which pre-processing method most effectively mitigates these differences. Furthermore, we explored the impact of different pre-processing methods on feature selection and the predictive power of radiomics models.

# 2. Materials and Methods

## 2.1 Dataset

The publicly available multicentre I-SPY1 TRIAL dataset is comprised of dynamic contrast-enhanced (DCE) MRI scans and tissue-based biomarkers, which were utilized previously to predict pathological complete response (pCR) and relapse-free survival (RFS) [1], [11], [14]. The scans used in this study were acquired on 3 different 1.5 Tesla scanners and the imaging protocols encompassed sagittal dynamic contrast-enhanced T1-weighted gradient echo sequences with specific parameters, such as TE = 4.5 ms, TR ≤ 20 ms, 16-18 cm field of view, flip angle ≤ 45º, minimum matrix 256x192, slice thickness ≤ 2.5 mm, and 64 slices [14]. ROIs were automatically generated using a U-net model and manually edited and approved by a board-certified, fellowship-trained breast radiologist.

Additionally, the PARTNER dataset comprises patients recruited in a randomized, phase 2/3, 3 stage trial to evaluate the safety and efficacy of the addition of Olaparib to platinum-based neoadjuvant chemotherapy in breast cancer patients with Triple Negative Breast Cancer (TNBC) and/or germline BRCA mutation. All scans were performed as standard-of-care breast MRI scans at Cambridge University Hospitals NHS Foundation Trust, Addenbrooke's Hospital including DCE MRI. The scans were acquired on a 1.5 Tesla MRI scanner (MR750; GE Healthcare) and DCE was obtained with a minimum matrix of 512x512 and a slice thickness of ~ 2mm and up to 98 slices. Images of the first phase post contrast injection were used for ROI drawing and extraction of radiomic features. ROIs were manually drawn by a board-certified radiologist (R.W.) specialised in breast imaging with more than ten years of experience in breast imaging.

161 subjects from the I-SPY1 TRIAL and 43 subjects from the PARTNER dataset were included in this study.

| Dataset | Flip angle (°) | Age (Y) | TR (s) | TE (s) | TI (s) | FOV (cm) | Manufacturer |
|---|---|---|---|---|---|---|---|
| ISPY1 | 10-55 | 27-68 | 2-4 | 1-6 | 0-30 | 18-30 | Philips Medical Systems (n=12) GE Medical Systems (n=110) Siemens (n=40) |
| PARTNER | 10 | 24-74 | 2 | 2-4 | 14-17 | 34-40 | GE Healthcare (n=43) |

*Table 1: Acquisition parameters for the two datasets*

## 2.2 MRI pre-processing

In this paper, we performed four common intensity normalization (IN) techniques used in MRI radiomics analysis, as described below. In addition, we applied spatial normalization (SN) using the SimpleITK package in Python to rescale the 3D MRI data to a common isotropic voxel size of 1mm³ per voxel via linear interpolation, and we evaluate their impact on the robustness of radiomics features extracted from MRI scans against various normalization techniques. The goal is to identify the most appropriate normalization method for future MRI radiomics studies, but, more importantly, to identify those radiomics features that are robust against using different normalization methods, which increases their significance for radiomics analysis across multiple different datasets and studies.

### 2.2.1 N4 Bias Field Correction

N4 bias field correction (BC) is a pre-processing algorithm used to correct for intensity inhomogeneities in MR images. Intensity inhomogeneities can arise due to variations in the magnetic field and can cause problems in quantitative analysis and segmentation of MR images. The BC algorithm estimates a smooth bias field and divides the original image by this estimated field to remove the inhomogeneity [29]. The N4 algorithm uses a non-parametric approach to estimate the bias field. It first fits a B-spline to the image intensities to capture the low-frequency components of the bias field. Then, a gaussian mixture model is used to estimate the high-frequency components of the bias field. These two components are combined to estimate the final bias field, which is then used to correct the original image. BC is a widely used algorithm due to its robustness to noise and other image artifacts. It is also computationally efficient and can be easily integrated into existing MR image analysis pipelines.

In this study, we use the N4BiasFieldCorrection function built into the SimpleITK package in Python [30] to apply BC on MRI images. The default parameters of this algorithm (4 resolution values, 50 iterations per level and a mask defined by Otsu thresholding [31], [32]) proved rather suboptimal as was noticed in the work in [33], since they provide a rather flat bias field estimation which does not sufficiently eliminate the inhomogeneities in the image. We tested Bias Field Correction using different masks as input for the algorithm, as well as modifying the number of iterations per fitting level and number of fitting levels. Eventually we selected the same parameters that were used in [33] using a segmented mask of the breast with 5 fitting levels and 50 iterations per fitting level for correction, which provided the most homogeneous distribution of intensity across the image without obviously overfitting the estimated bias field to the given image.

The breast segmentation images were obtained using a simple thresholding approach. To perform such thresholding, we normalized the MRI intensity values, acquired a binary volume using Otsu thresholding and finally morphologically cleaned up the binary volume.

Normalizing the intensity values of the volume to values between 0 and 1 aids in standardizing the data, for the segmentation to work better across different images and datasets. By excluding the first and last percentiles, the normalization becomes more robust against outliers, ensuring that extreme values do not disproportionately affect the overall normalization. We assume an MRI scan denoted as $im$ and the normalized scan as $im_n$. The equation for normalizing the scan while excluding the first and last percentile is:

$$im_n = \frac{im - im_{min}}{im_{max} - im_{min}} \qquad (1)$$

$im_{min}$ and $im_{max}$ representing the values of the first and last percentiles, respectively.

The next step involves computing a binary image using Otsu-thresholding. Otsu-thresholding is a technique that automatically determines an optimal threshold value to separate the foreground (object of interest) from the background in an image. It selects a threshold that minimizes the intra-class variance of the pixel intensities, maximizing the separation between foreground and background. By computing a binary image, where foreground pixels are assigned the value 1 and background pixels are assigned the value 0, the resulting image presents a clear distinction between the object and its surroundings. To eliminate small remaining holes and imperfections in the segmentation, a morphological closing operation is applied using the scipy package in Python and only the largest connected component is kept in the final segmentation of the breast, which was visually inspected.

The effect of the Bias Field Correction can be seen in Figure 1.

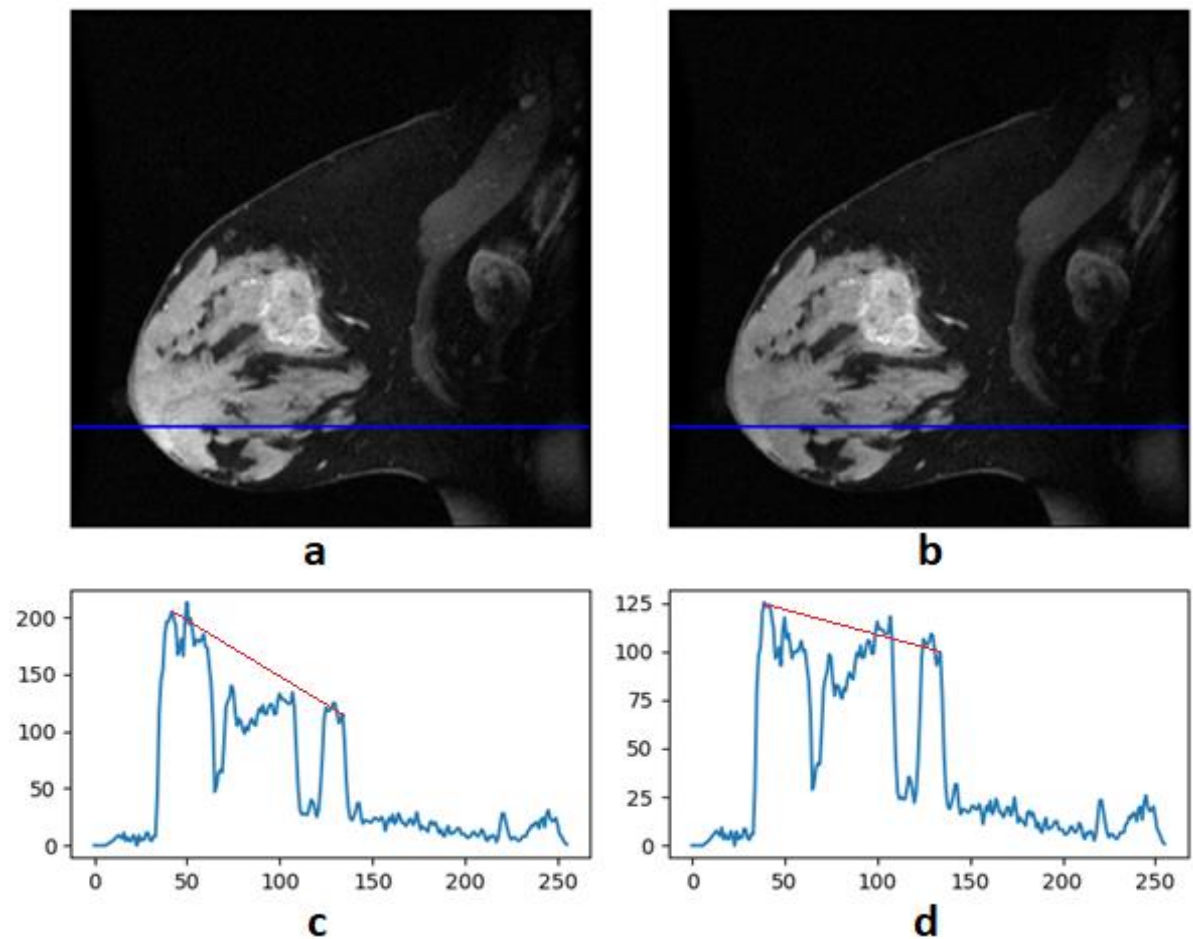

*Figure 1: Effect of Bias Field Correction on a sagittal slice of an example MR image of the ISPY1 dataset. The left column shows the uncorrected image (a) and intensity profile (c) along the blue line drawn on the image. The right column shows the corrected image (b) and intensity profile (d) along the blue line drawn on the image. After Bias Field correction (d) the slope of the intensity profile – as visualized by the red line – is significantly flatter than in the uncorrected image (c), indicating a more even distribution of intensities within the scan.*

### 2.2.2 Min-Max Normalization

Min-Max normalization, also known as feature scaling, is a simple and widely used technique for normalizing MRI scan intensities. The aim of this method is to transform the intensities so that they fall within a specified range, usually between 0 and 1, which standardizes the intensity range for easier comparison and analysis. To perform Min-Max normalization, the minimum and maximum intensities of the image are first determined. The image is then transformed by subtracting the minimum intensity and dividing by the range of intensities (i.e., the difference between the maximum and minimum values). This rescales the image intensities to fall within the range of 0 to 1. Min-Max normalization is particularly useful for ensuring consistency between MRI scans or subjects with varying intensity ranges, as it allows for a consistent range of values to be used for all images. It is also computationally efficient and easy to implement, making it a popular choice for many applications. The Min-Max normalized image $im_n$ is given by the equation

$$im_n(x,y,z) = \frac{im(x,y,z) - \min(im)}{\max(im) - \min(im)} \qquad (2)$$

for all voxel positions $(x, y, z)$ and $im$ representing the input image.

However, it is important to note that min-max normalization is highly sensitive to intensity outliers. These outliers can significantly impact the normalization process, potentially compressing the normalized intensity range. In other words, the presence of outliers can cause the range of intensities to be "squeezed" towards the middle of the scale, compromising the effectiveness of the normalization. One simple solution to this problem is the use of the very similar 1%-99% normalization which works by only taking the first and 99th percentile of the intensity values for normalization and crops the upper and lower values. Depending on the given data this may be the preferred choice. However, this was not feasible in our radiomics analysis since the intensity values of the lesion lie inside the highest percentile, which would have caused the normalization to crop values inside our ROI, not desirable for radiomics feature extraction (see Figure 2).

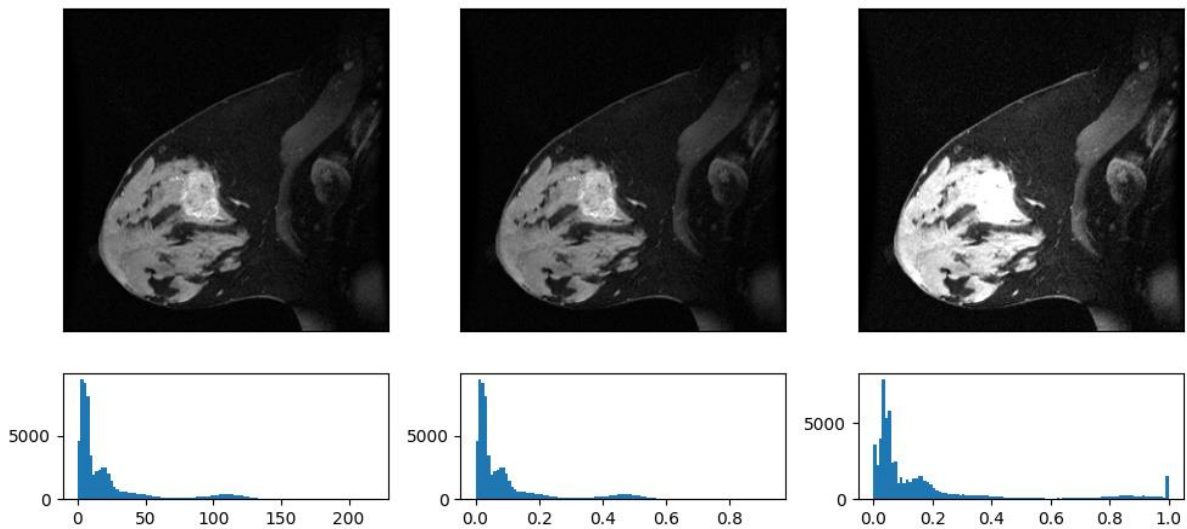

*Figure 2: Bias Field corrected sagittal slice of an example MR image of the ISPY1 dataset (top left), Min-Max normalized image (top centre) and 1%-99% normalized image (top right) with their respective histograms (bottom row). Min-Max normalization provides the same contrast and information as non-normalized image with a squeezed histogram, whereas 1%-99% normalization significantly changes the contrast. Fibroglandular and fatty tissue within the breast show higher contrast, but values within the lesion are completely saturated, as it can also be seen in the right border of the histogram where a peak appears near 1.0.*

### 2.2.3 Z-Score Normalization

Z-score normalization is a popular technique for standardizing MRI images across different subjects or scans. The aim of this method is to transform the image intensities so that they have a mean of zero and a standard deviation of one, making it easier to compare different images. To perform z-score normalization, the mean and standard deviation of the image intensities are calculated across the entire image or a specified ROI. Each voxel in the image is then transformed by subtracting from its intensity the mean and dividing divided by its standard deviation. This results in a new image where approximately 99.7% of all values fall within the range of 3 standard deviations, making the new intensity range fall roughly between -3 and 3, with most values around zero.

The basic equation for z-score normalization is

$$im_n(x, y, z) = \frac{im(x, y, z) - \mu}{\sigma} \tag{3}$$

with $im(x, y, z)$ representing the intensity value of the image at voxel position $(x, y, z)$ and $\mu$ and $\sigma$ representing the mean and standard deviation of the data respectively. We consider 3 different variations of z-score normalization, based on the region used to calculate the statistical parameters $\mu$ and $\sigma$. The normalized image, however, is always calculated for all voxels in the image, using equation 3. The different variations of z-score normalization used are as follows:

- Approach 1: Whole Image z-score normalization (ZA)
  This is the simplest approach to z-score normalization, as it uses all voxel positions in the image to calculate the values for $\mu$ and $\sigma$. This does not require any additional segmentation, as the entire image is used. However, the drawback of this approach is that it also includes unnecessary or even unwanted background tissue for calculation of the statistical parameters which could potentially skew the normalized result.
- Approach 2: Breast Tissue z-score normalization (ZB)
  This approach uses only the information contained in the breast tissue to eliminate the problem of including background information in the calculation of the statistical parameters $\mu$ and $\sigma$. To achieve this first a binary segmentation of the breast tissue in the MRI is required. The segmented image is calculated using the approach explained in section 2.2.1.
  Thus, the following equations are achieved:

$$\mu = \frac{1}{N} \sum_{(x,y,z) \in M_b} im(x, y, z) \tag{4}$$

and

$$\sigma = \sqrt{\frac{1}{N-1} \sum_{(x,y,z) \in M_b} (im(x, y, z) - \mu)^2} \tag{5}$$

With $M_b$ representing the set of the segmented coordinates and $N$ being the number of segmented pixels in $M_b$.

- Approach 3: Lesion Tissue z-score normalization (ZL)

  This approach is similar to the breast tissue z-score normalization as the region of the image that is used to calculate $\mu$ and $\sigma$ is limited. However, the difference is that only the region of the lesion itself is used. Therefore, equations (4) and (5) are modified to

$$\mu = \frac{1}{N} \sum_{(x,y,z) \in M_l} im(x, y, z) \qquad (6)$$

and

$$\sigma = \sqrt{\frac{1}{N-1} \sum_{(x,y,z) \in M_l} (im(x, y, z) - \mu)^2} \qquad (7)$$

with $M_l$ representing the set of the lesion segmentation coordinates and $N$ representing the number of segmented pixels in $M_l$. This lesion segmentation is used for radiomics feature extraction as denoted in section 2.3.

Figure 3 shows the effect of the 3 z-score normalization approaches on a given image. Since z-score normalization is a linear transformation the contrast of the different images stays the same, with only the mean and standard deviation changing. The effects of this normalization on mean and standard deviation can be observed in the histograms.

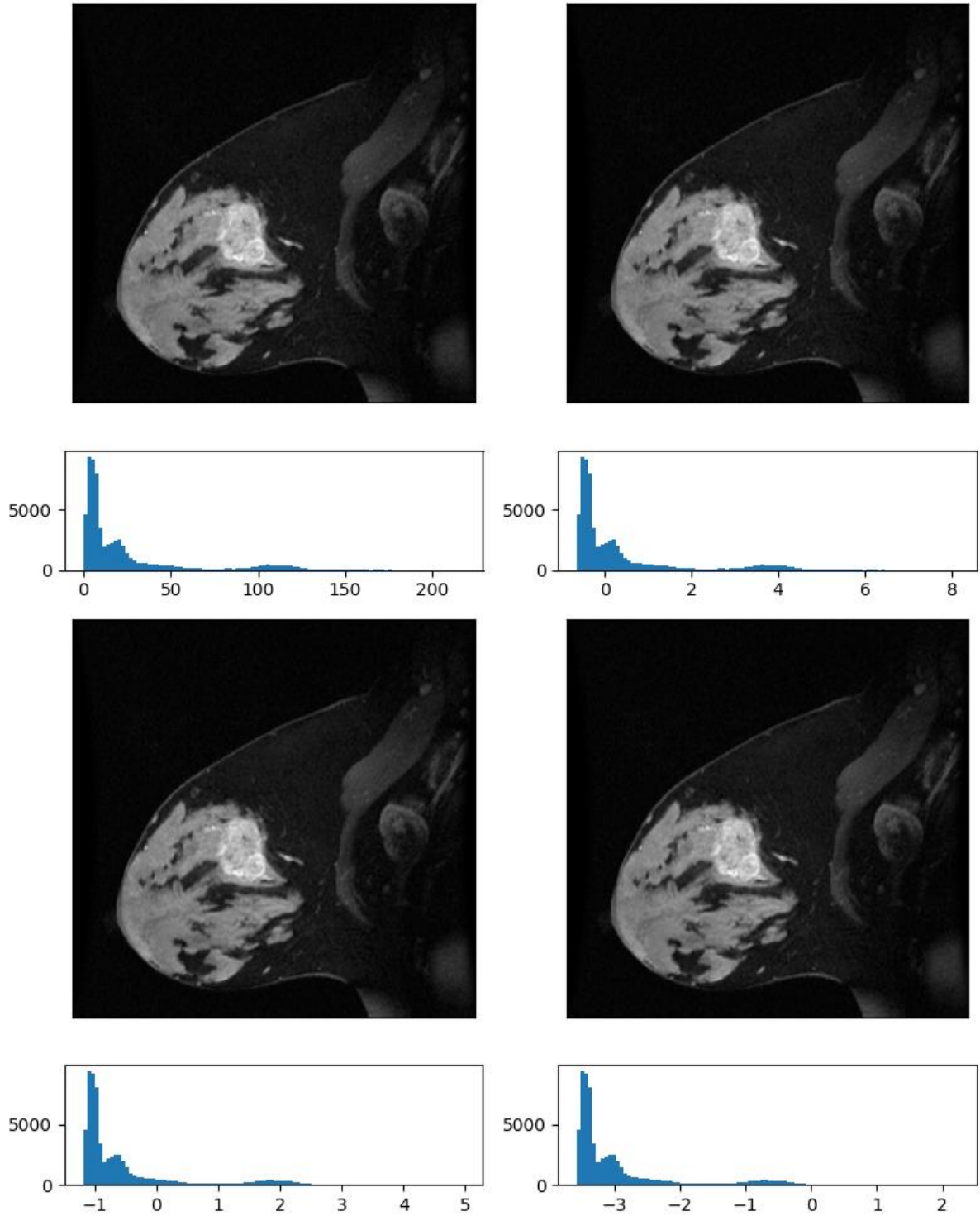

*Figure 3: Different variations of z-score normalization on the same MR image as in Figure 2. All z-score normalized images are based on the Bias Field corrected image (top left), with z-score over all (top right), z-score over breast (bottom left) and z-score over lesion (bottom right). The histograms below each image show the difference in the intensity level and range between the individual normalization approaches.*

### 2.2.4 Piecewise linear histogram equalization

Piecewise linear histogram equalization (PLHE) is a histogram-based normalization technique that is often used to enhance image contrast in MRI scans. This normalization technique can be applied on an image by dividing the image histogram into several segments and applying a linear stretching to each segment. This improves the distribution of image intensities, resulting in improved contrast and visibility of structures in the image. In contrast to the normalization techniques described in section 2.2.2 and 2.2.3, which are purely linear transformations, PLHE is a non-linear normalization technique, as it can change the relative differences between neighbouring voxels. The division of the image histogram into segments can be done manually or using an automatic binning algorithm. For each segment, a linear transformation is applied to stretch the intensities so that they cover a wider range. The stretched segments are then combined to produce the final equalized image. Piecewise linear histogram equalization is often preferred over traditional histogram equalization, as it can better preserve the original image contrast and avoid over-enhancement of the noise. It is also computationally efficient and easy to implement, making it a popular choice for enhancing image contrast in a variety of image modalities [34], [35].

Similar to [33] we calculate a normalized reference histogram based on a random subset of our MRI scans using the methods detailed in [21], [22]. To perform PLHE, we first divide the image histogram into several segments, with each segment containing an equal number of pixels. Then, for each segment, we calculate the cumulative distribution function (CDF) of the pixel intensities within that segment. We use this CDF to map the original voxel intensities in the segment to a new set of voxel intensities that are evenly spaced across the range of intensities in the segment. Finally, we combine the voxel intensities from all segments to create the equalized image. The effect of our PLHE implementation can be seen in

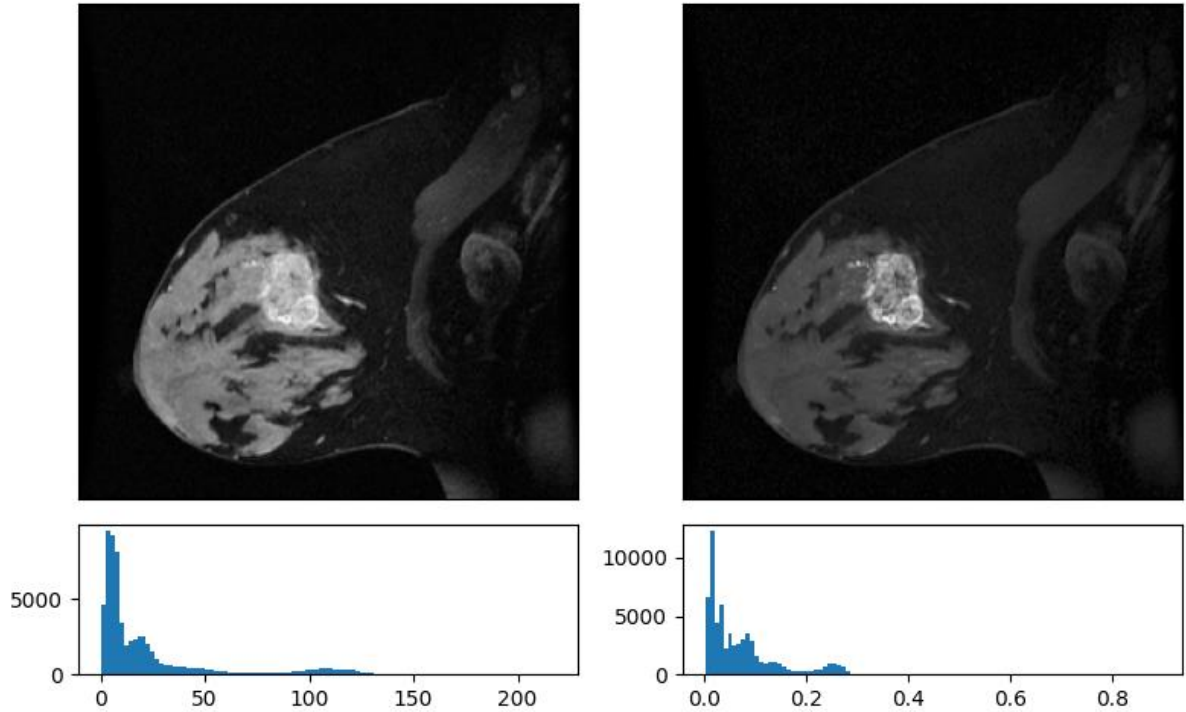

Figure 4.

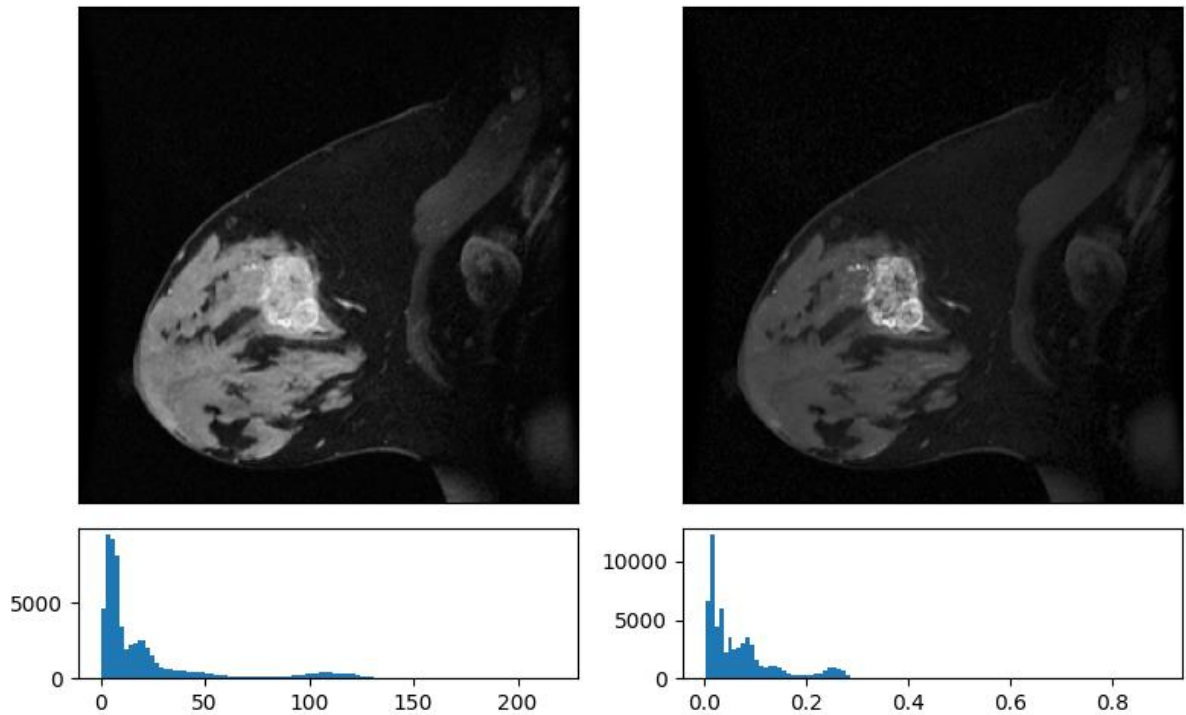

*Figure 4: Bias Field corrected image (top left) and Bias Field + PLHE normalized image (top right) with their respective histograms (bottom row).*

## 2.3 Radiomics Feature Extraction

We used the Pyradiomics toolkit (version 3.0.1) [36], a Python-based open-source software package, to compute a total of 107 features in the following categories:

Shape features pertain to the geometric attributes of a lesion, comprising its size, volume, and compactness amongst others. On the other hand, first-order features calculate the statistical properties of voxel intensities within the lesion, encompassing measures such as mean, variance, skewness, and kurtosis. Higher-order features, also referred to as texture features, explore the spatial relationships among groups of voxels within lesions. Texture features are quantified through various methods or techniques, such as gray-level co-occurrence matrices (GLCM), gray-level run-length matrices (GLRLM), gray-level size zone matrix (GLSZM), neighbouring gray tone difference matrix (NGTDM), and gray-level dependence matrix (GLDM).

In our research, we utilized the Pyradiomics toolkit (version 3.0.1), a Python-based open-source software package, to extract radiomics features from MRI data [36]. Pyradiomics offers a comprehensive array of standard feature extraction techniques specifically designed for medical images. The feature extraction configuration employed in this study incorporated all feature classes without filters, resulting in a total of 107 features.

Radiomic features were extracted using the following settings: Input image type was set to "original", meaning that none of the internal pre-processing methods were applied prior to feature extraction, thus ensuring that only the image normalization methods discussed above were applied on the images. Moreover, we established the bin count at 100 bins, ensuring an appropriate level of granularity for the analysis. For feature extraction a single volume ROI for the largest lesion was considered per scan.

## 2.4 Robustness Calculation

ICC calculation was performed using the Pingouin package version 0.5.3 – an open-source statistical package for Python [37]. The value of ICC can range from 0 to 1, with ICC values less than 0.5 typically being considered poor agreement, values between 0.5 and 0.75 moderate agreement, values between 0.75 and 0.9 good agreement, and values above 0.9 being considered excellent agreement [38]. In our study we are interested in the robustness of radiomics features against various image normalization techniques. Each image normalization combination can be interpreted as a separate grader with the radiomics feature value being the graded value. Thus, we have a fixed set of k graders grading every single subject exactly once, making the ICC3 model the most reasonable choice.

## 2.5 Evaluation of homogeneity of normalization techniques

Before performing a homogeneity analysis, it is necessary to investigate and account for any potential systematic differences among the scans to be compared. This includes differences and biases in the population of the two cohorts, but also systematic differences arising from the acquisition of the scans. To accomplish this, we focus our analysis on only one breast cancer subtype (TNBC) and examine whether there are any systematic age differences among patients across the various scanners and studies.

We categorized the data into 4 groups - one for each scanner. We applied the Shapiro test to determine if the age distribution is normal within each group. Upon confirming normality, we proceeded with an analysis of variance (ANOVA) test to rule out any significant age differences among the groups. To assess comprehensively the homogeneity of the radiomics features across scanners, given that only some of the features exhibit a normal distribution, we employ the Kruskal-Wallis test, which is well-suited for non-normally distributed data. This test allows us to compare all features and evaluate if there are statistically significant differences among the groups. Given that shape-based features are independent of any intensity normalization applied and change only very slightly when applying spatial transformations, we excluded those features from our homogeneity analysis. By running the Kruskal-Wallis test for each feature across the split set, we obtain p-values that indicate the confidence with which we can reject (p<0.05) the null hypothesis. In this case, the null hypothesis represents the absence of a significant difference between the groups. Our primary objective is to identify the normalization technique that minimizes the number of features where significant differences exist between groups (=scanners). Thus, we perform this test for all features, excluding shape-based features, and apply it to each normalization method under consideration.

## 2.6 The impact of different normalization techniques on prediction performance of radiomics models

To evaluate the impact of different normalization techniques on the prediction performance of pCR, we trained machine learning models on the TNBC patients of the I-SPY1 dataset (38 cases) and tested them on the PARTNER dataset (43 cases). The pCR labels were binarized for both datasets as 0 for non-complete response and 1 for complete response. The number of cases with complete response was 14 (prevalence of 37%) and 22 (prevalence of 51%) in the I-SPY1 and PARTNER dataset, respectively. In our analysis, 24 types of normalization were performed on the training and test set. For each normalization type, radiomics features from the largest lesion in each patient were only used as in Ref. [39]. As mentioned in Section 2.3, a total of $n = 107$ features were extracted. Within a stratified 5-fold cross validation (CV) loop, a correlation coefficient matrix $C \in \mathbb{R}^{n \times n}$, was constructed using the absolute spearman-r measure. Highly correlated features $C_{i,j>i} > r$, are typically outcasted as irrelevant and removed from the feature pool. Instead, they are given a second chance based on their relevance for classification as in Ref. [39]. The feature relevance is dictated by univariate Logistic Regression, wherein features are used individually to predict the response. A union of features thus selected from all CV folds, and the uncorrelated features $C_{i,j>i} < r$, is taken to form the final pool of features, which can typically fall between $20 - 80\%$ of n, depending on the dataset. Note that $r$ is a hyperparameter that should be optimized. The model hyperparameters were then tuned in a stratified 5-fold cross validation loop to select the best hyperparameters that yield the best Receiver Operating Characteristic – Area Under the Curve (ROC-AUC) score. Finally, feature selection algorithms such as F-Score, Relief, Mutual

Information (MI), Gini Importance, LASSO, Genetic Algorithm (GA), Sequential Backward Search (SBS), Sequential Forward Search (SFS), and Recursive Feature Elimination (RFE) were also employed to select the best features on the training set.

We used the Logistic Regression (LR) classifier with L1 and L2 regularization, commonly known as ElasticNet, with balanced class weights and the SAGA solver for the training. To tackle the issue of class imbalance in the training set (I-SPY1), we aided the model by using class weights proportional to the inverse class frequencies for the loss function, $W_c = N/(2 \times N_c)$, where $N_c$ is the number of samples in class $c$ and $N$ is the total number of samples. This implies a class weight of 0.79 and 1.36 for non-complete response and complete response, respectively. In the CV loop, each feature in the training fold was scaled using StandardScaler [40] and these statistics were used to further scale the validation fold. At the end of an iteration, the k$^{th}$ model was trained and evaluated on the validation fold using the evaluation metric. The average performance reported by the ensemble of k models measured on the corresponding validation folds is the CV-score. These models were then applied to the PARTNER dataset to test their generalisability using the selected features. The pre-processing and modelling pipelines were developed using the scikit-learn API version 1.0.2 [40]. Hyperparameters for LR that control the regularization such as l1_ratio and C were tuned within the CV loop [39] . Hyperparameter tuning was performed using the Optuna API version 3.0.0 [41] .

# 3. Results

## 3.1 Radiomics feature robustness

Radiomics feature robustness was estimated on a per feature basis between all 24 variations of normalization combinations. Shape-based features were excluded from calculation, as they are not affected by any IN techniques and only very slightly affected by SN. A subset of robustness estimation for 5 different normalization comparisons (no normalization compared to: SN; BC; BC+ZB; BC+PLHE respectively, in columns 1-4 and BC+ZB compared to BC+PLHE in column 5) is shown in Figure 5. First-order features are almost unaffected by SN alone but are less robust when applying BC. IN techniques can significantly affect first-order feature values, thus greatly reducing their robustness (Figure 5, row 1).

Higher order features, on the other hand, are more robust against mostly linear IN techniques, due to the relative changes in neighbouring voxel values changing only slightly. This can be seen by the relatively high ICC values – mostly >0.8 (Figure 5, rows 2-6, columns 1-3). Some combination, however, feature a high asymmetry of ICC distribution between the two datasets, which is especially noticeable for glszm features (Figure 5, row 4, columns 1-3), or ngtdm features (Figure 5, row 6, columns 4-5), indicating that the feature robustness also depends on the dataset itself.

Non-linear IN techniques, however, significantly reduce higher order feature robustness (Figure 5, columns 4-5). Especially, glszm and gldm features show a large distribution of ICC values, indicating an overall reduced robustness of radiomics features when comparing with radiomics features extracted from images normalized using non-linear IN techniques.

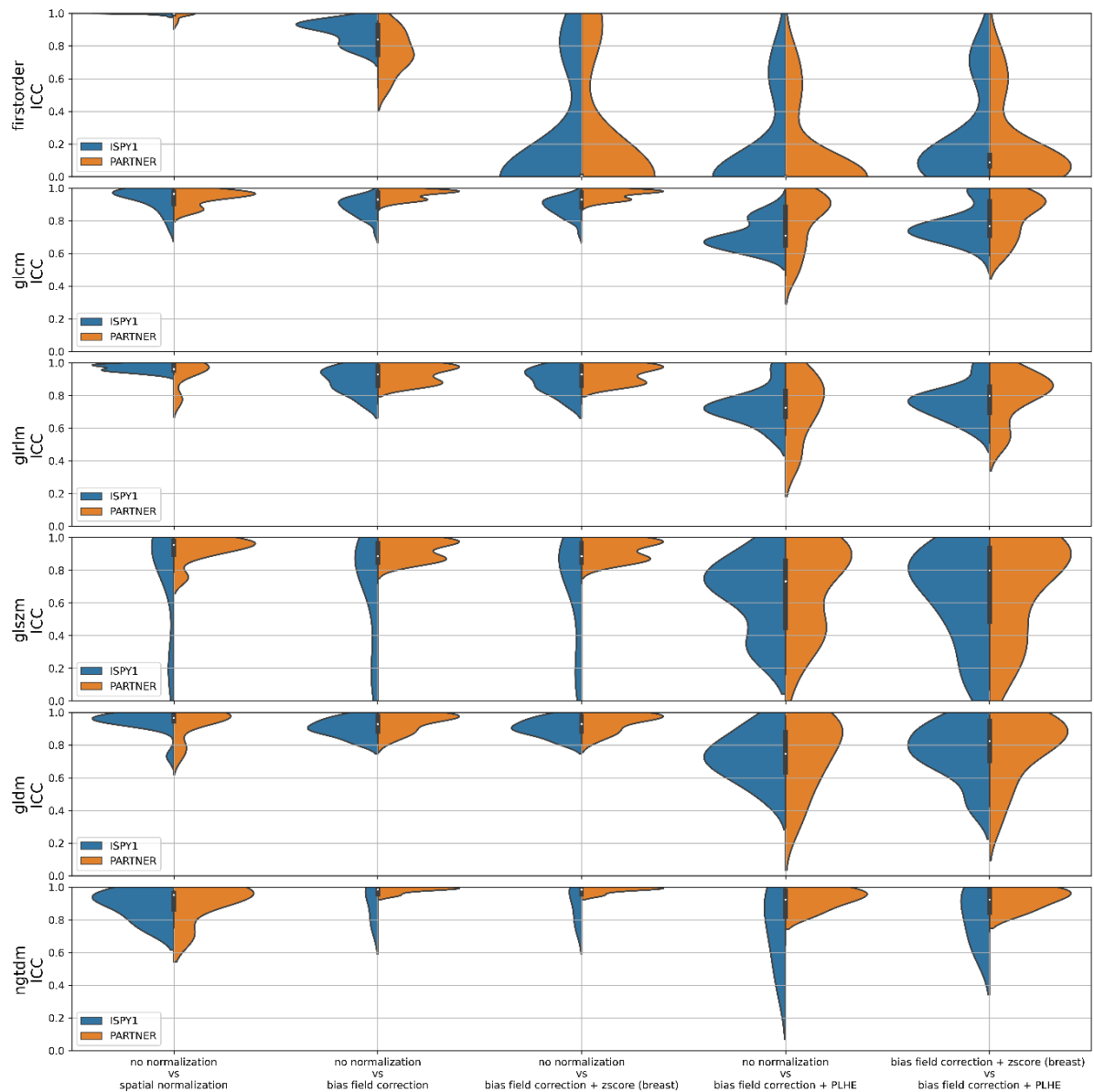

*Figure 5: Violin plot showing the ICC distributions for different normalization approaches. Every row contains the ICC values for a specific radiomics feature category. The columns denote the comparison of the normalization approaches, which are in order from left to right: no normalization vs only resampling; no normalization vs BC alone; no normalization vs BC + ZB; no normalization vs BC + PLHE; BC + ZB vs BC + PLHE. Every violin is split vertically to show the ICC distribution for the ISPY1 dataset (left) and PARTNER dataset (right).*

GLSZM features show a very large asymmetry for the first three normalization comparisons with ISPY1 ICC values covering the entire range between 0 and 1, whereas PARTNER data was consistently above 0.75. One reason for this may be the heterogeneity of the ISPY1 dataset (due to it being a multicentre study) whereas the PARTNER data set was acquired in a single centre. Radiomics feature robustness was spread particularly widely for comparisons involving PLHE. This might be explained by PLHE being a piecewise linear approach that can change relative differences between voxels while all other examined intensity normalization methods only scale the image linearly.

For the analysis of feature robustness over all normalization combinations instead of specific comparisons, Figure 6 shows an error bar plot of the ICC per feature, including the 95% confidence interval drawn with whiskers. First-order features are not robust against all different normalization methods and they are very comparable between the two datasets except for Entropy, Kurtosis, Skewness and Uniformity which have a much higher robustness in the PARTNER dataset than in the ISPY1 set (Figure 6). A very clear difference between the two sets can also be seen when examining the higher order features. Where ISPY1 has a rather scattered robustness across the features, the PARTNER dataset almost exclusively features robustness of at least 0.75 with only few features being the exception. The top performing features with ICC>0.9 in both datasets are provided in Table 2.

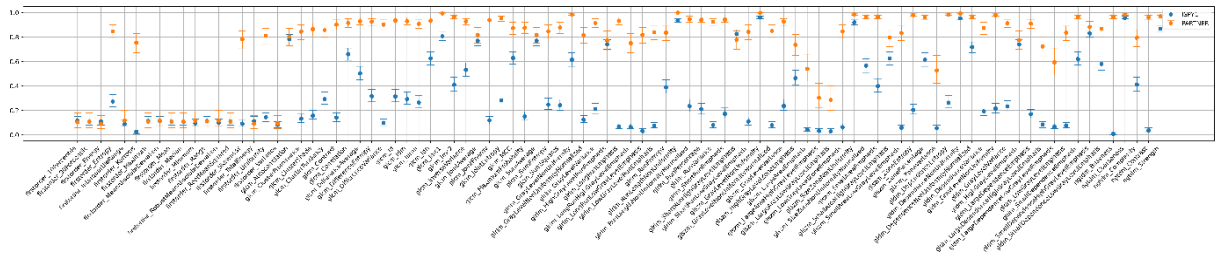

*Figure 6: ICC robustness by radiomics feature, including 95% confidence intervals drawn with whiskers for ISPY1 dataset (blue) and PARTNER dataset (orange).*

| Feature name | ICC (ISPY1) | ICC (PARTNER) |
|---|---|---|
| Glrlm_RunLengthNonUniformity | 0.94 | 0.99 |
| Glszm_GrayLevelNonUniformity | 0.96 | 0.99 |
| Glszm_SizeZoneNonUniformity | 0.92 | 0.99 |
| Gldm_DependenceNonUniformity | 0.95 | 0.99 |
| Ngtdm_Coarsness | 0.96 | 0.98 |

*Table 2: Features with excellent robustness across multiple normalization methods in both datasets*

Of the five features with excellent ICC > 0.9 (Table 2), four are related to non-uniformity measures, giving a measure of how much intensity values (or the derived matrices) vary within a ROI. Also, the 5$^{th}$ selected feature indicates the difference between voxels and their direct neighbours. Thus, regardless of the normalization approach used, the measure of non-uniformity in different variants remains mostly unaffected.

## 3.2 Feature homogeneity across different MR scanner manufacturers

The p-value distribution of the Kruskal-Wallis test for a set of normalization methods, applied on both datasets, is shown in Figure 7, with the red line indicating the threshold of 0.05 for significant differences. A clear trend can be seen where 8 out of 18 first-order features were not significantly different between scanners for 3 or more different normalization approaches. Looking at texture features, however, proportionally fewer features showed homogeneity across scanners when using multiple normalization methods. Some texture features showed very strong homogeneity for the following normalization methods: GLCM_JointEnergy showed p-values of 0.95 and 0.98 for bias field corrected PLHE in original and rescaled resolution, respectively, and p-values of 0.55 and 0.53 for the non-bias field corrected respective PLHE variants, whereas all other normalization combinations led to significant differences between scanners with p-values below 0.05. For other features like GLDM_GrayLevelNonUniformity,

GLDM_LargeDependenceLowGrayLevelEmphasis or NGTDM_Busyness the choice of normalization had no effect on their p-values below 0.05, indicating that the systematic difference between the data cannot be reduced with normalization alone for those features. Overall, without any normalization there were 8 radiomics features that did not show significant difference between the devices (Figure 8, first column). For bias field correction or resampling there was no clear trend in increasing or decreasing numbers of homogeneous features, but instead it depends on the specific combination of which intensity normalization, bias field correction and spatial normalization is applied. Table 3 provides a detailed overview of how many features were not showing significant differences between groups for all normalization combinations. Overall PLHE was the best performing intensity normalization technique in terms of homogenizing the MRI data across different scanners. Where the other features achieved somewhere between 3 and 14 homogeneous features, PLHE yielded 16 and more homogeneous features. The highest performing combination was PLHE with bias field correction and original voxel spacing reaching 34 homogeneous features.

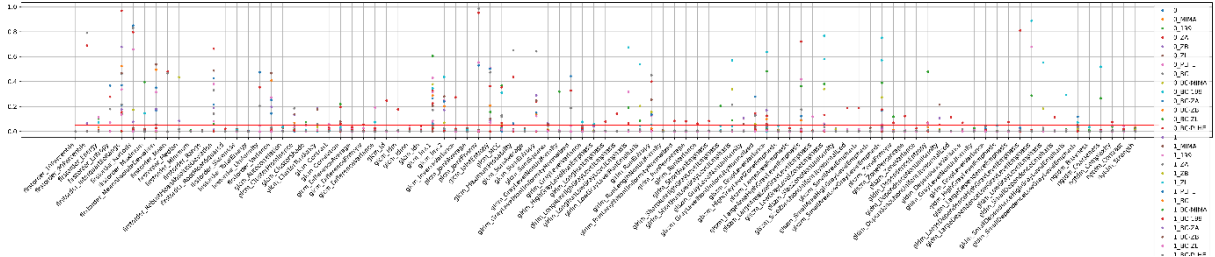

*Figure 7: p-value Distribution over the individual radiomics features. Red line indicates the cut-off value of 0.05 for rejecting the null-hypothesis that there is no significant difference between the groups.*

| Intensity normalization | Number of homogeneous features | | | |
|---|---|---|---|---|
| | **Without Bias field correction** | | **With bias field correction** | |
| | **Original spacing** | **Resampled** | **Original spacing** | **Resampled** |
| **None** | 8 | 9 | 6 | 3 |
| **Min-Max** | 8 | 10 | 6 | 3 |
| **1%-99%** | 14 | 8 | 13 | 6 |
| **ZA** | 13 | 8 | 10 | 7 |
| **ZB** | 7 | 12 | 10 | 7 |
| **ZL** | 4 | 11 | 8 | 5 |
| **PLHE** | 16 | 20 | 34 | 16 |

*Table 3: Amount of features without significant differences between groups per normalization method.*

A more detailed representation of homogeneous features can be seen in Figure 8. The features that are homogeneous between scanners for most normalization methods are glcm_Imc1, glcm_JointEntropy, firstorder_InterquartileRange and firstorder_RobustMeanAbsoluteDeviation, staying homogeneous in 20, 18, 13 and 12 normalization approaches respectively.

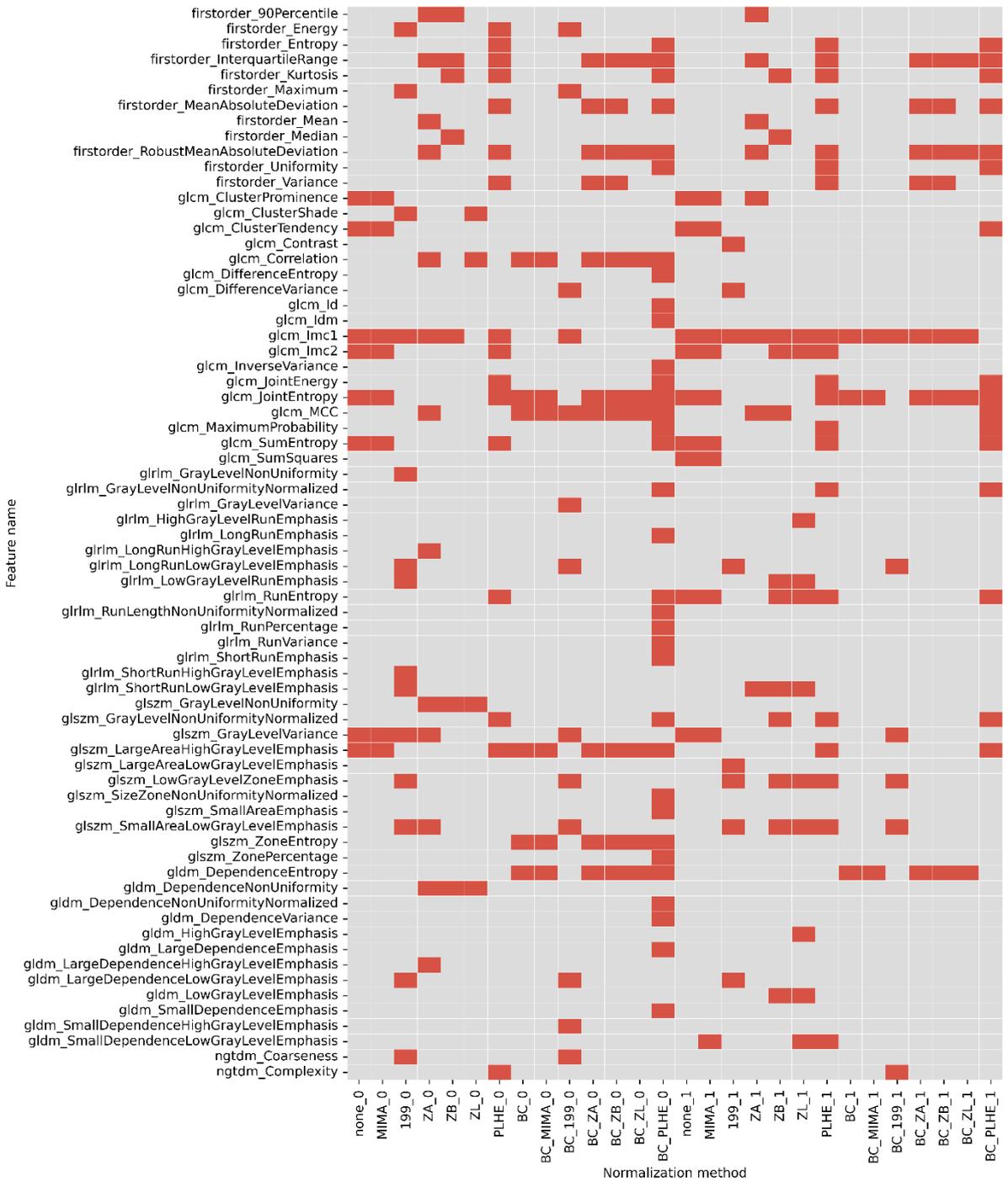

*Figure 8: Matrix plot showing homogeneous features across different scanner types for each normalization method shown as red squares. Y-Axis shows complete list of homogeneous features, X-Axis shows the normalization method. Labels on X-axis describe applied pre-processing steps separated via _. MIMA: min/max norm.; 199: 1%-99% norm.; ZA, ZB, ZL: z-score norm. overall, breast, lesion respectively; PLHE: piecewise linear histogram equalization; BC: bias field correction; 0/1: original resolution / isotropic resampling to 1x1x1mm per voxel*

## 3.3 Prediction performance of radiomic-based models under different normalization methods

The prediction results with the best performing feature selection algorithm, for all normalization techniques, for both the train and test datasets are given in Table 4. We observed that different feature selection algorithms performed best when different normalization methods were used confirming the influence of normalization on the feature selection process. We additionally report the percentage of improvement achieved on the test dataset by each normalization technique compared to no normalization. We observed a large difference (up to 18% improvement) between the prediction performance of the radiomic-based model when applying normalization compared to the case with no normalization. Without any intensity normalization, the models and feature selection algorithms do not converge on both datasets (AUC of 0.5), regardless of whether bias field correction (BC) or spatial normalization (SN) was applied.

The best-performing combination is with PLHE+BC, but without SN, resulting in an AUC score of ~0.68 on the test dataset. This is in correspondence with the high homogeneity measure reported in Section 3.2 for PLHE normalization. The z-score normalization based on breast tissue (ZB+BC), without SN, gives a similar AUC score on the test set but with less generalization (AUC on test data set dropped around 17% compared to train dataset).

| Intensity normalization (IN) | Bias field correction (BC) | Spatial normalization (SN) | FS Algorithm | Features selected | CV score on I-SPY1 | Test score on PARTNER | Improvement compared to no-normalization |
|---|---|---|---|---|---|---|---|
| **None** | ✗ | ✗ | F-Score | 1 | 0.5 (0.0) | 0.5000 (0.0000) | 0% |
|  |  | ✓ | F-Score | 1 | 0.5 (0.0) | 0.5000 (0.0000) | 0% |
|  | ✓ | ✗ | F-Score | 8 | 0.56 (0.1479) | 0.5662 (0.0371) | 7% |
|  |  | ✓ | F-Score | 1 | 0.5 (0.0) | 0.5000 (0.0000) | 0% |
| **Min-Max** | ✗ | ✗ | RFE | 9 | 0.9267 (0.076) | 0.5801 (0.0227) | 8% |
|  |  | ✓ | F-Score | 1 | 0.7133 (0.182) | 0.5294 (0.0165) | 3% |
|  | ✓ | ✗ | F-Score | 1 | 0.7567 (0.1942) | 0.5195 (0.0000) | 2% |
|  |  | ✓ | F-Score | 1 | 0.8133 (0.1726) | 0.5281 (0.0000) | 3% |
| **PLHE** | ✗ | ✗ | SFS | 5 | 0.8133 (0.1193) | 0.6450 (0.0153) | 15% |
|  |  | ✓ | F-Score | 1 | 0.7067 (0.1623) | 0.5883 (0.0494) | 9% |
|  | ✓ | ✗ | **SFS** | **15** | **0.7667 (0.2211)** | **0.6788 (0.0246)** | **18%** |
|  |  | ✓ | F-Score | 3 | 0.8033 (0.1574) | 0.5887 (0.0000) | 9% |
| **ZA** | ✗ | ✗ | F-Score | 10 | 0.7667 (0.0667) | 0.6472 (0.0195) | 15% |
|  |  | ✓ | RFE | 9 | 0.81 (0.0925) | 0.5147 (0.0277) | 1% |
|  | ✓ | ✗ | F-Score | 9 | 0.7967 (0.2136) | 0.5861 (0.0054) | 9% |
|  |  | ✓ | F-Score | 1 | 0.8033 (0.107) | 0.5931 (0.0000) | 9% |
| **ZB** | ✗ | ✗ | MI | 5 | 0.7533 (0.1693) | 0.6316 (0.0112) | 13% |
|  |  | ✓ | Gini | 1 | 0.7733 (0.265) | 0.4654 (0.0000) | -3% |
|  | ✓ | ✗ | RFE | 8 | 0.85 (0.1708) | 0.6814 (0.0329) | 18% |
|  |  | ✓ | F-Score | 3 | 0.79 (0.2033) | 0.6000 (0.0108) | 10% |
| **ZL** | ✗ | ✗ | MI | 8 | 0.7667 (0.1563) | 0.5238 (0.0395) | 2% |

| | | | | | | |
|---|---|---|---|---|---|---|
| | | ✓ | RFE | 10 | 0.8167 (0.1708) | 0.4372 (0.0204) | -6% |
| | ✓ | ✗ | Relief | 3 | 0.6933 (0.2153) | 0.5918 (0.0128) | 9% |
| | | ✓ | SBS | 7 | 0.7 (0.2646) | 0.5671 (0.0307) | 7% |

*Table 4: Prediction results for all variations of normalization techniques. The performance on the training and testing dataset is reported, grouped by the intensity normalization method, the absence (✗) and presence (✓) of bias field correction or spatial normalization. The best performing combination of normalizations is highlighted in bold.*

Features selected for each normalization type are represented in Figure 9. Features such as shape_Elongation, firstorder_Minimum, glcm_Imc2, glcm_MCC were the most frequently selected features across all normalization types. It is known that shape features are not affected by intensity normalization techniques and are naturally robust, whereas the remaining set of features do not belong to the group of features with excellent ICC scores (cf. Table 2).

Of the 15 features that were selected by the best reported normalization combination (PLHE+BC without SN), only a single feature, glcm_Imc1, had an ICC > 0.8 in both datasets. One third of the selected features had an ICC < 0.6 in the train dataset (firstorder_Kurtosis, glcm_Contrast, glcm_ClusterProminence, glcm_DifferenceEntropy, glcm_MCC), whereas those same features had an ICC > 0.8 in the test dataset. This behaviour agrees with the results reported in Section 3.1 and associated Figure 6, that suggest there are systematic differences between the train and test datasets. Additionally, only one third of the selected features appear in the high homogeneity group (cf. Figure 7 and Figure 8: firstorder_InterquartileRange, firstorder_MeanAbsoluteDeviation, firstorder_Kurtosis, glcm_DifferenceEntropy, glcm_MCC). These results suggest that robustness or homogeneity alone does not translate to high predictive power or generalization to unseen datasets.

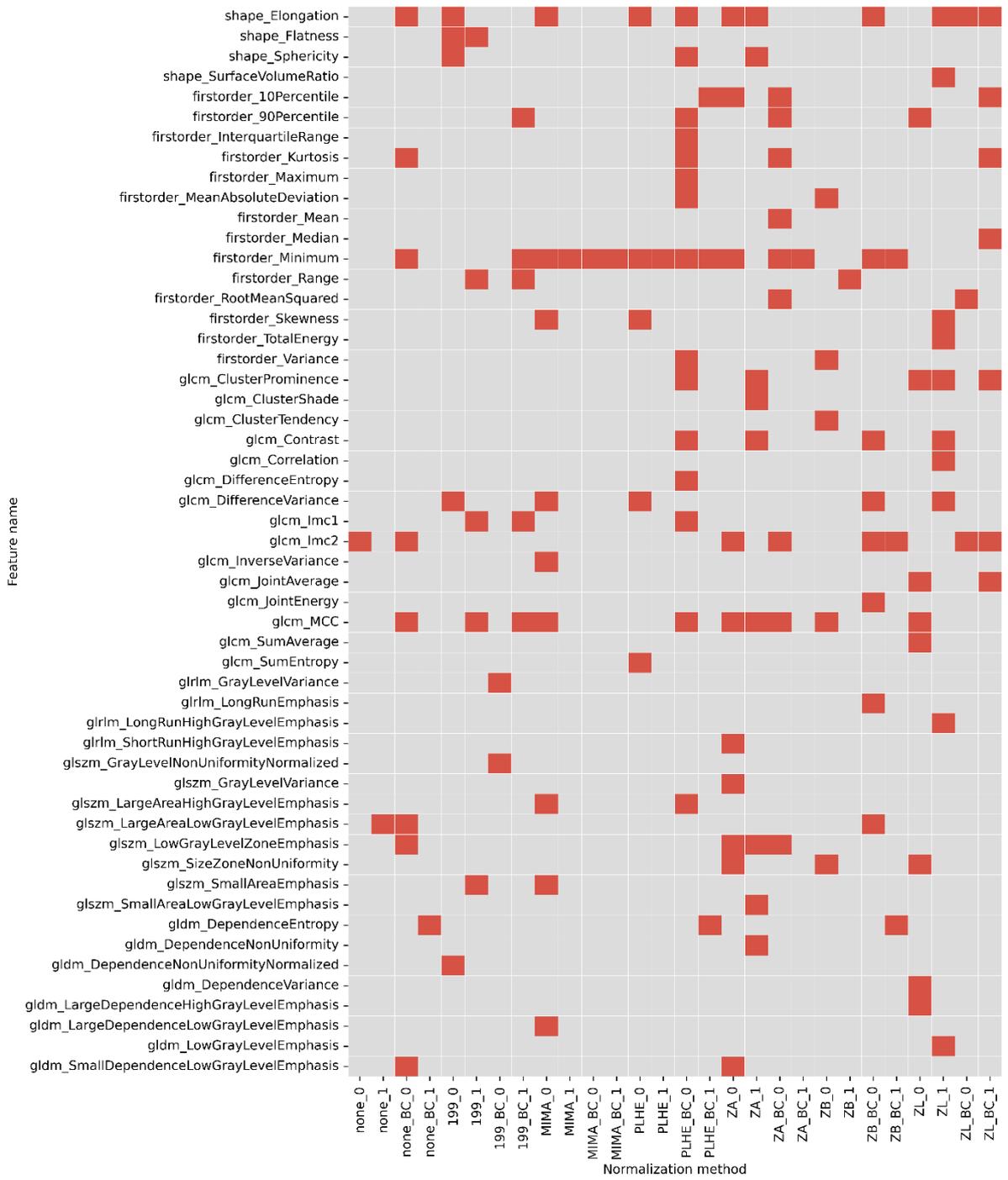

*Figure 9: Matrix plot showing the optimal selected features for each normalization method as red squares. Y-Axis shows complete list of selected features, X-Axis shows the normalization methods. Labels on X-axis describe applied pre-processing steps separated via _. MIMA: min/max norm.; 199: 1%-99% norm.; ZA, ZB, ZL: z-score norm. overall, breast, lesion respectively; PLHE: piecewise linear histogram equalization; BC: bias field correction; 0/1: original resolution / isotropic resampling to 1x1x1mm per voxel*

# 4. Discussion

The study presented here evaluated the effects of different normalization methods, alone and in combination, on radiomics features calculated for breast cancer on contrast-enhanced MRI. We provide insight into the effect that acquiring images with different scanner models has on the numerical values of radiomic features, analysing whether such effect can be reduced with various normalisation techniques, and how these different normalisations might in turn have an impact on the feature selection and predictive modelling. Importantly, our results demonstrate that robustness or homogeneity alone does not necessarily translate into high predictive power or generalization to unseen datasets. Therefore, strictly enforcing the selection of only the most robust features may negatively impact model performance. In this study, we show that features with moderate ICC, even if not exceptionally robust, can still significantly enhance classification performance. This suggests that features with a balance of robustness and predictive relevance can be more beneficial for model performance than prioritizing robustness alone.

We have also shown that the choice of normalization method strongly affects the robustness of radiomics feature categories and individual features, and that the size of this effect varies not only between features and categories but also between datasets. Except for most first-order features, the ICC for all different normalization methods was much higher for the PARTNER than the ISPY1 data set (Figure 5, Figure 6). One possible explanation is the high heterogeneity of the ISPY1 data set in terms of image acquisition whereas all scans of the PARTNER data set were acquired on a single scanner. Likely the normalisation methods used in this study cannot fully compensate for these differences in terms of data heterogeneity. Of note, the low robustness of first-order features can be explained with different normalization methods normalizing to different intensity levels, thus making first-order features not directly comparable between certain normalization methods. Radiomic features of non-uniformity on the other hand remain largely unaffected (and thus robust) when calculated following different image normalization methods (Table 2), but appear to be strongly affected by the scanner type used (Figure 7). Therefore, the normalization methods reported here are unlikely to be able to account for scanner differences for these features.

As the main objective of the study was at the preprocessing level, to identify robust normalization techniques that work consistently across different datasets or acquisition scenarios, the lack of generalization cannot be simply attributed to model overfitting. This suggests that the choice of normalization approach can have a significant effect on the predictive power and generalisability (Table 3). Without any intensity normalization the model predictions are mostly random, with no significant improvement when applying bias field correction and/or spatial resampling. A simple min-max normalization, however, already improves the predictive capability by 8% with further improvements when using PLHE and z-score approaches except for the lesion-only based z-score normalization. PLHE, ZA and ZB increased the predictive power even more without any additional bias field correction – showing that intensity normalization is essential when working with multiple MRI datasets. When alternating bias field correction and spatial normalization with these three intensity normalization methods, in general, bias field correction further improved the predictive power, whereas spatial normalization had a negative effect. Since bias field correction homogenizes the individual images, the positive effect in combination with another intensity normalization approach is quite self-explanatory. In contrast, spatial normalization negatively affected predictive power likely because it leads to a reduced spatial resolution for a large proportion of scans and thus also reduces image information.

# 5. Conclusion

In this study, we investigated the impact of different MRI normalization techniques on the robustness and stability of radiomics features. Additionally, we examined how the choice of normalization technique affects the generalization of a predictive model to an external dataset. Our findings demonstrate that systematic differences between MRI scanners can significantly affect many radiomics features across two data sets and scanner models. However, we discovered that the combination of PLHE with bias field correction can mitigate this effect on a larger number of features compared to other normalization methods tested in this paper. Furthermore, we observed that the robustness of features is not solely dependent on the normalization method, but also influenced by specific characteristics of the dataset itself. For instance, variations in patient demographics, imaging protocols, or tumour heterogeneity between datasets can alter the stability and predictive power of radiomics features. This highlights the importance of considering both factors when conducting radiomics analyses. For the prediction of complete pathological response, we found that applying either z-score normalization on the breast tissue or PLHE – both in combination with bias field correction, yielded the best performance. Nevertheless, it is worth noting that the robustness of radiomics features between different normalization techniques varied significantly, with features selected in the two best performing methods (ZB and PLHE) having very poor robustness. This emphasizes the need for standardization of normalization techniques in radiomics analysis of breast MRI scans. Overall, our study provides valuable insights into the effects of MRI normalization techniques on radiomics features and the generalization of predictive models. Our findings highlight the importance of carefully selecting and standardizing normalization methods to ensure accurate and reliable radiomics analyses for breast MRI scans.

# 6. Acknowledgement

We thank the PARTNER Trial Consortium for their contribution (see supplementary materials).

# Bibliography

[1] K. Clark *et al.*, 'The Cancer Imaging Archive (TCIA): maintaining and operating a public information repository', *J. Digit. Imaging*, vol. 26, pp. 1045–1057, 2013.

[2] Abraham, J.E., Pinilla, K., Dayimu, A. *et al.* The PARTNER trial of neoadjuvant olaparib in triple-negative breast cancer. *Nature* (2024). https://doi.org/10.1038/s41586-024-07384-2

[3] J. D. Shur *et al.*, 'Radiomics in Oncology: A Practical Guide', *RadioGraphics*, vol. 41, no. 6, pp. 1717–1732, Oct. 2021, doi: 10.1148/rg.2021210037.

[4] C. Scapicchio, M. Gabelloni, A. Barucci, D. Cioni, L. Saba, and E. Neri, 'A deep look into radiomics', *Radiol. Med. (Torino)*, vol. 126, no. 10, pp. 1296–1311, Oct. 2021, doi: 10.1007/s11547-021-01389-x.

[5] J. E. van Timmeren, D. Cester, S. Tanadini-Lang, H. Alkadhi, and B. Baessler, 'Radiomics in medical imaging—"how-to" guide and critical reflection', *Insights Imaging*, vol. 11, no. 1, p. 91, Aug. 2020, doi: 10.1186/s13244-020-00887-2.

[6] R. Da-ano *et al.*, 'Performance comparison of modified ComBat for harmonization of radiomic features for multicenter studies', *Sci. Rep.*, vol. 10, no. 1, Art. no. 1, Jun. 2020, doi: 10.1038/s41598-020-66110-w.

[7] F. Orlhac *et al.*, 'A Postreconstruction Harmonization Method for Multicenter Radiomic Studies in PET', *J. Nucl. Med.*, vol. 59, no. 8, pp. 1321–1328, Aug. 2018, doi: 10.2967/jnumed.117.199935.

[8] A. Ibrahim *et al.*, 'Radiomics for precision medicine: Current challenges, future prospects, and the proposal of a new framework', *Methods*, vol. 188, pp. 20–29, Apr. 2021, doi: 10.1016/j.ymeth.2020.05.022.

[9] M. E. Mayerhoefer *et al.*, 'Introduction to Radiomics', *J. Nucl. Med.*, vol. 61, no. 4, pp. 488–495, Apr. 2020, doi: 10.2967/jnumed.118.222893.

[10] R. T. Shinohara *et al.*, 'Statistical normalization techniques for magnetic resonance imaging', *NeuroImage Clin.*, vol. 6, pp. 9–19, 2014.

[11] N. M. Hylton *et al.*, 'Neoadjuvant chemotherapy for breast cancer: functional tumor volume by MR imaging predicts recurrence-free survival—results from the ACRIN 6657/CALGB 150007 I-SPY 1 TRIAL', *Radiology*, vol. 279, no. 1, pp. 44–55, 2016.

[12] A. Crombé *et al.*, 'Intensity harmonization techniques influence radiomics features and radiomics-based predictions in sarcoma patients', *Sci. Rep.*, vol. 10, no. 1, Art. no. 1, Sep. 2020, doi: 10.1038/s41598-020-72535-0.

[13] H. Mi *et al.*, 'Impact of different scanners and acquisition parameters on robustness of MR radiomics features based on women's cervix', *Sci. Rep.*, vol. 10, no. 1, Art. no. 1, Nov. 2020, doi: 10.1038/s41598-020-76989-0.

[14] D. Newitt, N. Hylton, and others, 'Multi-center breast DCE-MRI data and segmentations from patients in the I-SPY 1/ACRIN 6657 trials', *Cancer Imaging Arch*, vol. 10, no. 7, 2016.

[15] M. Hatt, M. Vallieres, D. Visvikis, and A. Zwanenburg, 'IBSI: an international community radiomics standardization initiative', *J. Nucl. Med.*, vol. 59, no. supplement 1, pp. 287–287, May 2018.

[16] B. D. Wichtmann *et al.*, 'Influence of Image Processing on Radiomic Features From Magnetic Resonance Imaging', *Invest. Radiol.*, vol. 58, no. 3, p. 199, Mar. 2023, doi: 10.1097/RLI.0000000000000921.

[17]  P. Chirra et al., *Empirical evaluation of cross-site reproducibility in radiomic features for characterizing prostate MRI*. 2018. doi: 10.1117/12.2293992.

[18]  M. Shafiq-ul-Hassan et al., 'Intrinsic dependencies of CT radiomic features on voxel size and number of gray levels', *Med. Phys.*, vol. 44, no. 3, pp. 1050–1062, Mar. 2017, doi: 10.1002/mp.12123.

[19]  M. Bologna, V. Corino, and L. Mainardi, 'Technical Note: Virtual phantom analyses for preprocessing evaluation and detection of a robust feature set for MRI-radiomics of the brain', *Med. Phys.*, vol. 46, no. 11, pp. 5116–5123, 2019, doi: 10.1002/mp.13834.

[20]  A. Carré et al., 'Standardization of brain MR images across machines and protocols: bridging the gap for MRI-based radiomics', *Sci. Rep.*, vol. 10, no. 1, Art. no. 1, Jul. 2020, doi: 10.1038/s41598-020-69298-z.

[21]  L. G. Nyúl and J. K. Udupa, 'On standardizing the MR image intensity scale', *Magn. Reson. Med. Off. J. Int. Soc. Magn. Reson. Med.*, vol. 42, no. 6, pp. 1072–1081, 1999.

[22]  M. Shah et al., 'Evaluating intensity normalization on MRIs of human brain with multiple sclerosis', *Med. Image Anal.*, vol. 15, no. 2, pp. 267–282, 2011.

[23]  S. G. K. Patro and K. K. Sahu, 'Normalization: A Preprocessing Stage', *IARJSET*, pp. 20–22, Mar. 2015, doi: 10.17148/IARJSET.2015.2305.

[24]  Y. Li, S. Ammari, C. Balleyguier, N. Lassau, and E. Chouzenoux, 'Impact of Preprocessing and Harmonization Methods on the Removal of Scanner Effects in Brain MRI Radiomic Features', *Cancers*, vol. 13, no. 12, Art. no. 12, Jan. 2021, doi: 10.3390/cancers13123000.

[25]  J. C. Reinhold, B. E. Dewey, A. Carass, and J. L. Prince, 'Evaluating the Impact of Intensity Normalization on MR Image Synthesis', *Proc. SPIE-- Int. Soc. Opt. Eng.*, vol. 10949, p. 109493H, Mar. 2019, doi: 10.1117/12.2513089.

[26]  J. Panic, A. Defeudis, G. Balestra, V. Giannini, and S. Rosati, 'Normalization Strategies in Multi-Center Radiomics Abdominal MRI: Systematic Review and Meta-Analyses', *IEEE Open J. Eng. Med. Biol.*, vol. 4, pp. 67–76, 2023, doi: 10.1109/OJEMB.2023.3271455.

[27]  J. Antunes et al., 'Radiomics Analysis on FLT-PET/MRI for Characterization of Early Treatment Response in Renal Cell Carcinoma: A Proof-of-Concept Study', *Transl. Oncol.*, vol. 9, no. 2, pp. 155–162, Apr. 2016, doi: 10.1016/j.tranon.2016.01.008.

[28]  H. Moradmand, S. M. R. Aghamiri, and R. Ghaderi, 'Impact of image preprocessing methods on reproducibility of radiomic features in multimodal magnetic resonance imaging in glioblastoma', *J. Appl. Clin. Med. Phys.*, vol. 21, no. 1, pp. 179–190, 2020, doi: 10.1002/acm2.12795.

[29]  J. Juntu, J. Sijbers, D. Dyck, and J. Gielen, 'Bias Field Correction for MRI Images', in *Computer Recognition Systems*, M. Kurzyński, E. Puchała, M. Woźniak, and A. Żołnierek, Eds., in Advances in Soft Computing, vol. 30. Berlin, Heidelberg: Springer Berlin Heidelberg, 2005, pp. 543–551. doi: 10.1007/3-540-32390-2_64.

[30]  N. J. Tustison et al., 'N4ITK: improved N3 bias correction', *IEEE Trans. Med. Imaging*, vol. 29, no. 6, pp. 1310–1320, 2010.

[31]  N. Otsu, 'A Tlreshold Selection Method from Gray-Level Histograms'.

[32]  X. Xu, S. Xu, L. Jin, and E. Song, 'Characteristic analysis of Otsu threshold and its applications', *Pattern Recognit. Lett.*, vol. 32, no. 7, pp. 956–961, May 2011, doi: 10.1016/j.patrec.2011.01.021.

[33]  M.-J. S. Martin et al., 'A radiomics pipeline dedicated to Breast MRI: validation on a multi-scanner phantom study', *Magn. Reson. Mater. Phys. Biol. Med.*, vol. 34, pp. 355–366, 2021.

[34] S. Javadi, M. Dahl, and M. I. Pettersson, 'Adjustable Contrast Enhancement Using Fast Piecewise Linear Histogram Equalization', in *Proceedings of the 2020 3rd International Conference on Image and Graphics Processing*, Singapore Singapore: ACM, Feb. 2020, pp. 57–61. doi: 10.1145/3383812.3383830.

[35] W. R. Abdul-Adheem, 'ENHANCEMENT OF MAGNETIC RESONANCE IMAGES THROUGH PIECEWISE LINEAR HISTOGRAM EQUALIZATION', vol. 15, 2020.

[36] J. J. M. van Griethuysen *et al.*, 'Computational Radiomics System to Decode the Radiographic Phenotype', *Cancer Res.*, vol. 77, no. 21, pp. e104–e107, Oct. 2017, doi: 10.1158/0008-5472.CAN-17-0339.

[37] R. Vallat, 'Pingouin: statistics in Python', *J. Open Source Softw.*, vol. 3, no. 31, p. 1026, Nov. 2018, doi: 10.21105/joss.01026.

[38] T. K. Koo and M. Y. Li, 'A Guideline of Selecting and Reporting Intraclass Correlation Coefficients for Reliability Research', *J. Chiropr. Med.*, vol. 15, no. 2, pp. 155–163, Jun. 2016, doi: 10.1016/j.jcm.2016.02.012.

[39] S. Hatamikia, G. George, F. Schwarzhans, A. Mahbod, and R. Woitek, 'Breast MRI radiomics and radiomics-based predictions of response to neoadjuvant chemotherapy - how are they affected by variations in tumour delineation?', *Submitt. To*.

[40] F. Pedregosa *et al.*, 'Scikit-learn: Machine Learning in Python', *Mach. Learn. PYTHON*.

[41] T. Akiba, S. Sano, T. Yanase, T. Ohta, and M. Koyama, 'Optuna: A Next-generation Hyperparameter Optimization Framework'. arXiv, Jul. 25, 2019. Accessed: Jun. 19, 2023. [Online]. Available: http://arxiv.org/abs/1907.10902